\documentclass[english,aps,prc,0superscriptaddress,longbibliography,notitlepage]{revtex4-1}
\usepackage{amsmath,amssymb,multirow,bm}
\usepackage{soul}
\usepackage{epsfig}
\usepackage{graphicx}
\usepackage{amsmath}
\usepackage{lipsum}
\usepackage{color}
\makeatother

\usepackage{babel}
\usepackage[colorlinks,linkcolor=blue,anchorcolor=blue,citecolor=blue,urlcolor=blue,filecolor=black]{hyperref}

\newcommand{\beq}{\begin{equation}}
\newcommand{\eeq}{\end{equation}}
\newcommand{\bea}{\begin{eqnarray}}
\newcommand{\eea}{\end{eqnarray}}
\usepackage{graphicx}
\preprint{}

\begin{document}
\title{Neutron skins probed in proton knockout from neutron-rich nuclei}
\author{C.A. Bertulani}
\affiliation{Department of Physics and Astronomy, East Texas A\&M University, Commerce, TX 75428, USA}
\email{carlos.bertulani@etamu.edu}

\author{R.V. Lobato}
\affiliation{Centro Brasileiro de Pesquisas F\'isicas, Rua Dr. Xavier Sigaud, 150 - Urca, Rio de Janeiro, Brazil}
\email{lobato@cbpf.br}

\date{\today}

\begin{abstract}
Proton-induced quasifree knockout reactions provide a powerful probe of nuclear single-particle structure and reaction dynamics in both stable and neutron-rich nuclei. In this work we develop a unified theoretical framework for the calculation of inclusive (p,2p) and sequential (p,3p) reaction cross sections and fragment momentum distributions at intermediate and relativistic energies. The approach is based on a probabilistic extension of Glauber multiple-scattering theory combined with microscopic nuclear densities obtained from Hartree-Fock-Bogoliubov calculations using Skyrme energy-density functionals. We focus in particular on the sensitivity of total cross sections and longitudinal momentum dispersions to neutron-skin thickness along isotopic chains. Our results indicate that both (p,2p) and (p,3p) reactions exhibit a systematic decrease of cross section and momentum width with increasing neutron excess, reflecting enhanced attenuation and surface bias induced by neutron skins. The effect is significantly stronger for two-proton removal, suggesting that (p,3p) reactions may offer enhanced sensitivity to isovector nuclear structure. These findings establish proton-induced knockout reactions as complementary hadronic probes of neutron skins and the density dependence of the nuclear symmetry energy.
\end{abstract}

\maketitle

\section {Introduction} 

Single-nucleon removal reactions offer a direct probe of nuclear single-particle structure. Electron-induced knockout, (e,e'p), is especially clean because initial- and final-state distortions for electrons are modest, and the electromagnetic interaction is precisely characterized. Large experimental programs in the past, spanning many targets, extracted accurate proton spectroscopic factors (occupancies) and rms radii of bound orbits \cite{RevModPhys.28.214,DonnellyWalecka1975,RevModPhys.56.461}. In recent years, (p,2p) reactions have been studied extensively in experiments involving inverse kinematics and radioactive ion beams, becoming a workhorse for spectroscopy of unstable systems using heavy-ion projectiles and proton targets~\cite{Kobayashi2008_p2p,Yasuda:2010,Miklukho2013,Pan16,pty011,Atar:2018,Dia18,ribeiro:2018:PRC,PhysRevLett.122.162503,Holl2019,PhysRevLett.122.072502,10.1093/ptep/ptaa109,Panin23,Jung2023,10.1093/ptep/ptae013,Aumann2024-review}. In these reactions, the distortions are much stronger than in (e,e'p), but with suitable kinematics one can determine the total angular momentum $j$ in addition to the orbital angular momentum $l$, a distinct advantage over (e,e'p) \cite{Kobayashi2008_p2p,Yasuda:2010,Miklukho2013,Pan16,Wakasa17,pty011,Atar:2018,Dia18,ribeiro:2018:PRC,PhysRevLett.122.162503,Holl2019,PhysRevLett.122.072502,10.1093/ptep/ptaa109,Panin23,Jung2023,10.1093/ptep/ptae013,Aumann2024-review}. A further benefit of proton-induced reactions is the symmetric treatment of proton and neutron removal. Because distortions are strong, extracted quantities can depend on the reaction model and parameter choices, underscoring the importance of systematic benchmark tests on well-characterized stable nuclei spanning a range of masses and bound orbitals \cite{Wakasa17,aumann:2013:PRC,AumBar21}. The goal of (p,pN) reactions is to further explore the shell model, which has long provided a powerful account of bulk nuclear behavior; yet how structure evolves with increasing isospin imbalance remains a central open issue in quantum many-body physics. Changes in shell structure stem from properties of the underlying nucleon-nucleon and many-body interactions, driving vigorous efforts on both theory and experiment, including new radioactive-beam facilities and innovative techniques.

Interest has increasingly turned to nuclei far from stability. Intermediate-energy facilities now provide diverse rare-isotope beams, enabling reaction studies that extend structural knowledge across a wider swath of the nuclide chart. (p,pN) reactions in inverse kinematics are central to this effort~\cite{Kobayashi2008_p2p,Yasuda:2010,Miklukho2013,Pan16,pty011,Atar:2018,Dia18,ribeiro:2018:PRC,PhysRevLett.122.162503,Holl2019,PhysRevLett.122.072502,10.1093/ptep/ptaa109,Panin23,Jung2023,10.1093/ptep/ptae013,Aumann2024-review}.

Two-nucleon removal has drawn strong interest because it accesses additional facets of nuclear structure, notably NN correlations \cite{PhysRevLett.109.202505}. Recent work employing $^{80}$Zn(p,3p) populated new states in $^{78}$Ni that were not reachable via one-nucleon knockout, underscoring the promise of (p,3p) reactions for spectroscopy at the limits of stability \cite{Taniuchi2019}. Two-proton knockout from projectile nuclei was also measured in neutron-rich systems at 250 MeV/nucleon in inverse kinematics \cite{PhysRevLett.125.012501}. The dependence of the measured (p,3p) events on the angular distribution of the three emitted protons was investigated. Comparisons of the data with different theoretical models support the dominance of a sequential process in which the target proton collides with a proton inside the projectile, followed by a subsequent collision with another proton, resulting in the removal of both protons from the nucleus. This interpretation highlights that multistep reaction dynamics, rather than direct correlated pair knockout alone, can play a dominant role in (p,3p) observables.

We present a theoretical framework for the calculation of total (p,3p) cross sections and the momentum distribution of the residual ($A-2$) nucleus, assuming it survives the collision, for projectile energies in the range from 100~MeV/nucleon to a few~GeV/nucleon incident on proton targets. Robust interpretation of (p,3p) observables requires a clear picture of the reaction dynamics: ejecting a spatially correlated proton pair probes different physics than stepwise removal of uncorrelated protons \cite{PhysRevLett.125.012501}. For instance, when the two protons are removed independently, the recoil (residual-nucleus) momentum distribution primarily reflects the intrinsic structure of the residue; by contrast, short-range correlated pp removal ties the response to the pairs momentum distribution, which displays only a weak dependence on the mass number $A$. This distinction underscores the importance of disentangling reaction-induced effects from genuine many-body correlations. 

The neutron-skin thickness of a nucleus,
\begin{equation}
\Delta R_{np} = \sqrt{\langle r^2\rangle_n} - \sqrt{\langle r^2\rangle_p},
\end{equation}
measures the extent to which neutrons extend beyond protons in the nuclear surface. Although small in absolute magnitude, typically $\Delta R_{np}\sim 0.1$--$0.3$~fm for heavy nuclei, the neutron skin encodes essential information about the nuclear equation of state (EoS), nuclear reactions, collective modes, and astrophysical phenomena such as neutron-star structure. The neutron skin is primarily determined by the density dependence of the symmetry energy $S(\rho)$ around sub-saturation densities $\rho\sim(0.5$--$0.7)\rho_0$. This density interval reflects the surface region of finite nuclei and represents an effective average sensitivity over sub-saturation densities rather than a probe at a single density. The pressure difference between neutrons and protons in the nuclear surface depends strongly on the slope parameter $L$ of the symmetry energy,
\begin{equation}
L = 3\rho_0 \left.{dS(\rho)}/{d\rho}\right|_{\rho_0}.
\end{equation}
A larger $L$ implies stronger neutron pressure and thus a thicker neutron skin, whereas a softer symmetry energy yields smaller neutron skins. While $L$ provides the dominant contribution to the correlation with $\Delta R_{np}$, higher-order terms in the symmetry energy, such as curvature contributions, introduce systematic model-dependent scatter. Because neutron skins are among the few terrestrial observables directly sensitive to $S(\rho)$ at sub-saturation densities, they serve as essential constraints on the EoS for neutron-rich matter. This connection has been established in mean-field frameworks and density functional theory~\cite{Centelles2009,Suzuki2022,RocaMaza2015,Piekarewicz2012}.

The same symmetry energy governing neutron skins also controls the pressure of neutron-rich matter inside neutron stars. This leads to several important correlations~\cite{Lattimer2023,Oertel2017,Fattoyev2018}: (a) neutron-star radius: stiffer symmetry energy (larger $L$) produces larger radii for $1.4\,M_\odot$ stars; (b) crust thickness and structure: neutron skins correlate with the crust-core transition density, nuclear pasta phases, and proton fractions in the outer core; (c) tidal deformabilities: the neutron-skin thickness influences the tidal response measured in gravitational-wave events. These correlations reflect robust trends across classes of EoS models rather than exact one-to-one mappings. Thus, accurate neutron-skin measurements provide direct constraints on neutron-star structure and gravitational-wave astrophysics \cite{tsang:2012:PRC,Horowitz2014}.

Fragmentation and knockout reactions, such as (p,2p) (and possibly (p,3p)), at intermediate and high energies could provide a potentially sensitive probe of neutron skins \cite{Kox1987,Zenihiro2018}. In these reactions, initial- and final-state interactions depend mainly on the total density $\rho=\rho_p+\rho_n$ and are strongly surface weighted due to absorption and attenuation. Sensitivity to neutron densities is enhanced because $\sigma_{pn} > \sigma_{pp}$ at several hundred MeV to GeV energies, making reaction dynamics increasingly neutron dominated. Glauber-type attenuation, nuclear transparency, and multiple-scattering effects therefore depend primarily on the surface matter distribution rather than on volume-averaged densities. This neutron dominance is most pronounced in the intermediate-energy regime where Glauber and eikonal descriptions are applicable and tends to saturate at asymptotically higher energies. Consequently, incorrect neutron densities can bias extracted spectroscopic factors, nuclear radii, and reaction cross sections by 5--20\% or more, primarily through reaction filtering and absorption rather than through bound-state normalization. Glauber and DWIA studies have shown that reaction observables such as $(p,2p)$ cross sections, nuclear transparency, and knockout spectroscopic factors require accurate neutron densities to be interpreted reliably~\cite{Hufner1985,Giusti1999,Tostevin1999,Hen2017,Ogata2015}. While proton-induced knockout reactions probe the nuclear interior, strong attenuation in the entrance and exit channels leads to an
effective surface weighting of the reaction amplitude, particularly for neutron-rich nuclei. This is consistent with previous studies showing that $(p,2p)$ reactions retain sensitivity to the interior
while exhibiting enhanced sensitivity to surface and density-gradient effects in asymmetric systems \cite{aumann:2013:PRC,Bertulani.104.L061602}.

Momentum distributions are particularly sensitive to the spatial extent of nucleon wave functions, and therefore can reflect surface properties such as neutron skins and halos.  In neutron-rich nuclei, valence neutrons may occupy weakly bound orbitals extending into the low-density surface region. The increased spatial localization length leads, via the uncertainty principle, to narrower momentum distributions. As a result, knockout momentum distributions provide a clean probe of density tails and surface diffuseness, complementing information obtained from interaction cross sections or elastic scattering. At higher missing momenta, typically above $p \sim 250$--$300~\mathrm{MeV}/c$, momentum distributions probe physics beyond the independent-particle picture. In this regime, contributions from short-range and tensor correlations dominate, giving rise to high-momentum tails that are absent in mean-field models. Measurements of these tails, and their isospin dependence in $(p,2p)$ versus $(p,pn)$ reactions, provide direct evidence for correlated nucleon pairs and help explain the systematic quenching of spectroscopic factors observed in knockout experiments \cite{Hen2017,subedi:2008:SCIENCE}.

Beyond their structural content, momentum distributions also serve as sensitive diagnostics of the reaction mechanism itself. Deviations in width, asymmetry, or energy dependence can signal the importance of multistep processes, core excitation, recoil effects, or limitations of the eikonal and Glauber approximations. For this reason, momentum distributions are often regarded as more reliable observables than absolute cross sections, whose normalization may depend strongly on reaction-model assumptions.

Ref. \cite{aumann:2013:PRC} has shown that experimentally observed momentum distributions are intrinsically reaction-dependent quantities, reflecting a nontrivial convolution of nuclear structure and reaction dynamics rather than a simple Fourier transform of the bound-state wave function. Reaction distortions further modify the shape of momentum distributions. Initial- and final-state interactions, recoil effects, and off-shell kinematics can lead to broadening, skewness, or asymmetries, even for nominally symmetric orbitals \cite{Giusti1999,aumann:2013:PRC,Ryckebusch2003}. An additional complication arises in the interpretation of high-momentum components. While short-range and tensor correlations in the nuclear wave function generate genuine high-momentum tails, similar features can also emerge from multistep scattering and rescattering processes inherent to the reaction mechanism. Without a consistent reaction model, it is therefore not possible to attribute high-momentum strength uniquely to many-body correlations. This ambiguity is particularly relevant for $(p,2p)$ and $(p,pn)$ reactions at GeV energies, where both correlated dynamics and reaction-induced effects may contribute \cite{Hen2017,subedi:2008:SCIENCE}.

These considerations also clarify the frequently observed dichotomy between the successful reproduction of momentum distribution shapes and the systematic quenching of absolute knockout cross sections. While spectroscopic factors are sensitive to missing correlations and many-body effects, the shapes of momentum distributions are dominated by reaction filtering and are therefore less affected by global normalization issues. Shape and normalization thus probe different aspects of nuclear dynamics and should not be conflated.

In this article we first outline a multistep reaction theory based on the Glauber method, a scheme shown to match experimental measurements with high fidelity in (p,2p) reactions \cite{aumann:2013:PRC}. We then use this framework for an in-depth study of key measurable quantities relevant to reactions with neutron-rich nuclides, emphasizing inverse-kinematics experiments that are expected to probe single-particle structure. We also report a formalism to obtain the longitudinal momentum dispersions of projectile-like fragments. We demonstrate a clear reduction of proton-induced knockout cross sections with increasing neutron excess, reflecting the surface bias of proton removal and the reduction of local Fermi momentum in low-density regions. These effects are entirely absent in simplified statistical descriptions such as the Goldhaber model \cite{Goldhaber1974,PhysRevC.39.460}, underscoring the importance of microscopic, density-dependent treatments. The hierarchy $\sigma^{(p,3p)}>\sigma^{(p,2p)}$ is naturally explained by the recoil mechanism and persists across the isotopic chain. The paper is organized as follows. In Sec.~\ref{sec:Xformalism} we outline the theoretical framework used in the calculations of (p,2p) and (p,3p) cross section. Section \ref{sec:Mformalism} presents a formalism for momentum distributions of fragments.  A detailed discussion of  numerical results and systematics of neutron-skin effects and their physical origin is given in each section. Conclusions and perspectives are summarized in Sec.~\ref{sec:conclusions}.

\section{Cross section formalism}
\label{sec:Xformalism}
 At energies of $\sim 200$ MeV and above the proton wavelength is small enough that semiclassical concepts can be used for the binary proton-proton collisions. There are two main methods to study such collisions. 
The Glauber multiple scattering theory  is based on the eikonal approximation and assumes that the projectile follows a straight-line trajectory through the nucleus, interacting with nucleons via elementary proton-nucleon scattering amplitudes. For example, the elastic scattering amplitude for a proton incident on a nucleus with mass number $A$ is written as
\begin{equation}
f_{pA}(\mathbf{q}) = \frac{i k}{2\pi} \int d^2\mathbf{b}\; e^{i \mathbf{q}\cdot \mathbf{b}}
\left\{ 1 - \left\langle \prod_{j=1}^{A} \left[ 1 - \Gamma_{pN}(\mathbf{b} - \mathbf{s}_j) \right] \right\rangle \right\}. \label{G0}
\end{equation}
where   $k$ is the incoming proton momentum in the  c.m. frame, $\mathbf{q}$ is the momentum transfer, $\mathbf{b}$ is the impact parameter (transverse coordinate),
and $\Gamma_{pN}$ is the profile function given by the Fourier transform of the elementary scattering amplitude
$\Gamma_{pN}(\mathbf{b}) = ({i\pi k}/2) \int d^2 \mathbf{q}e^{-i\mathbf{q}\cdot \mathbf{b}} f_{pN}(\mathbf{q})$ and is often parametrized as
\begin{equation}
\Gamma_{pN}(\mathbf{b}) = \frac{\sigma_{pN}^{\text{tot}} (1-i \alpha_{pN})}{4\pi \beta_{pN}} \;
\exp\!\left(- \frac{b^2}{2\beta_{pN}}\right). \label{G2}
\end{equation}
Also, $\mathbf{s}_j$ is the transverse coordinate of the $j$-th nucleon inside the nucleus, and $\langle \cdots \rangle$ in Eq. \eqref{G0} is the  average over the nuclear ground-state density distribution.
   $\sigma_{pN}^{\text{tot}}$ is the total proton-nucleon cross section, $\alpha_{pN}$ is the ratio of real to imaginary parts of the forward $pN$ scattering amplitude, and  $\beta_{pN}$ is the slope parameter. It determines the transverse range of the elementary proton-nucleon (pN) interaction. A finite $\beta_{pN}$  means the interaction has a Gaussian-like spread in impact parameter space.
Physically, this accounts for the fact that pN scattering is not point-like, but has a finite diffraction cone in momentum transfer  ${\bf q} = {\bf k}^\prime - {\bf k}$, with ${\bf k}^\prime$ (${\bf k}$) being the proton momentum in the lab (or c.m.) frame after (before) the collision. It is usual to allow $\alpha_{pN}$ and $\beta_{pN}$ to be free parameters to fit a multiple set of reaction cross sections. 
The total proton-nucleus reaction cross section can be written as
\begin{equation}
\sigma_{\text{R}} = \int d^2 b \; \left[ 1 - \left| S(\mathbf{b}) \right|^2 \right],  \ \ \ \ \ \ \ {\rm with} \ \ \ \ S(\mathbf{b}) = \left\langle \prod_{j=1}^{A} \left[1 - \Gamma_{pN}(\mathbf{b} - \mathbf{s}_j) \right] \right\rangle , \label{G1}
\end{equation}
being the $S$-matrix, or survival amplitude, at impact parameter $\mathbf{b}$.

The formalism described above, widely used in the analysis of proton-nuclear reactions at high energy collisions, complicates comparison with reaction cross section measurements  because it relies on fitting  ($\alpha, \beta$ and $\sigma$) to reproduce the data leading to a more difficult interpretation the results. A simpler approach is to neglect both the $\alpha$ as well as the $\beta_{pN}$ dependence  and concentrate on medium modifications of $\sigma_{pN}^{\text{tot}}$, as done in Refs. \cite{AumannPRL119,Teixeira22}. In this case, the S-matrix simplifies to
\begin{equation}
S(\mathbf{b}) =
\left\langle
\prod_{j=1}^{A}\left[
1 - \frac{\sigma_{pN}^{\text{tot}}}{2}\,
\delta^{(2)}(\mathbf{b}-\mathbf{s}_j)
\right]
\right\rangle . \label{G3}
\end{equation}
Another problem is that the ``straight-line approximation'', implicit in the Glauber method, Eqs. (\ref{G0}-\ref{G3}), is not adequate in the treatment of specific process and their details, such as in (p,2p) and (p,3p) reactions. 
The physical picture of the most commonly adopted Glauber formalism for proton-nucleus collisions is that the incoming proton travels straight through the nucleus at high energy and at each transverse position $\mathbf{b}$, the probability of interacting with nucleons is described by the product of survival factors $\left[1 - \Gamma_{pN}\right]$. Multiple scattering effects arise from the product over all nucleons: the proton can scatter off one, two, or more nucleons during its passage along a straight-line. Although this approximation neglects deflection and quantum interference effects, it has been shown to provide a reasonable leading-order description of inclusive observables at intermediate and high energies ($\gtrsim 200$ MeV/nucleon) \cite{Tostevin1999,AumannPRL119,Teixeira22}.

 \begin{figure}[t]
\begin{center}
\includegraphics[scale=0.42]{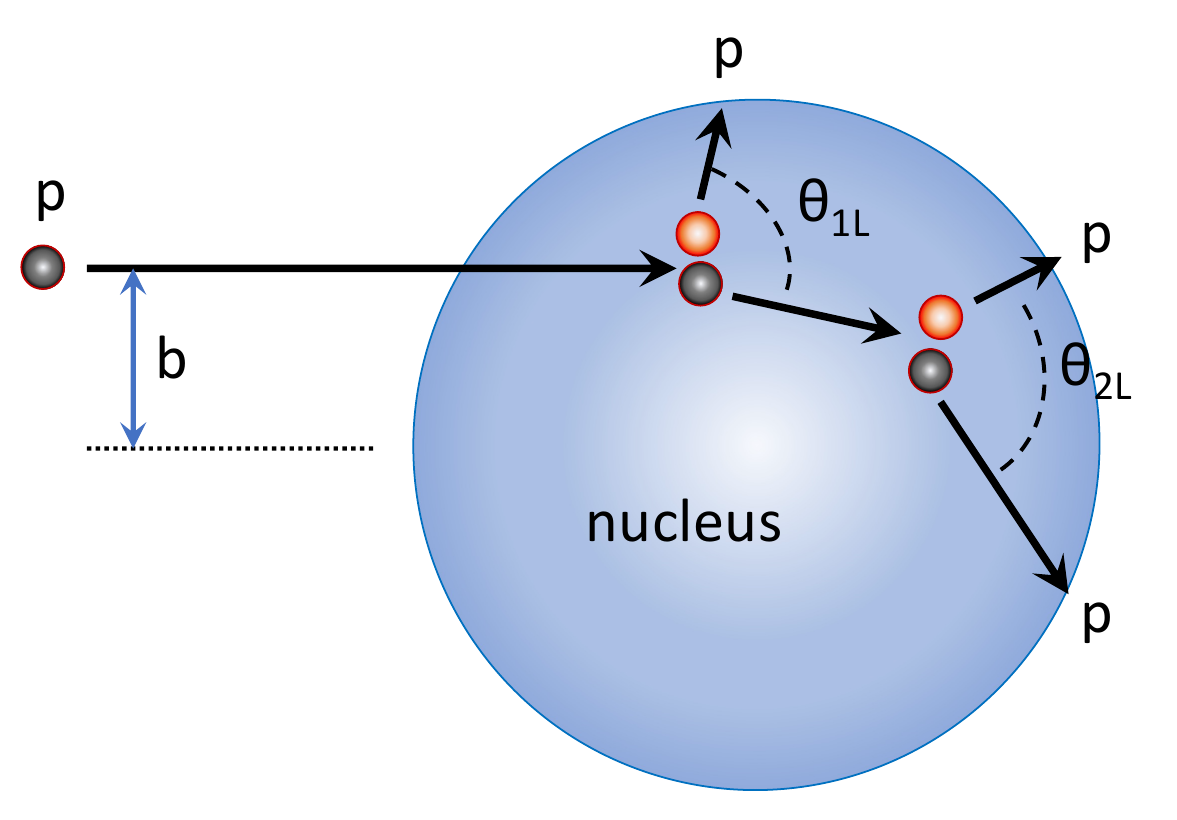}
\caption{ Schematic representation of a (p,3p) collision geometry considered in this work. The impact parameter ${\bf b}_1$, and ${\bf b}_1'$, correspond to the vector distance between the nuclear center  and the velocity vector for the leading proton, and the proton knocked out from the nucleus, after the first collisions. A similar notation, with ${\bf b}_2$  and ${\bf b}_2'$, is used for the second collision. \label{fig1}}
\end{center}
\end{figure} 

In this work we propose a refined extension of the multiple scattering formalism using eikonal scattering waves for the nucleon  propagation between  multiple binary collisions. Explicitly, the eikonal phase shift in the S-matrix, $S({\bf b}, z_1, z_2) = \exp\left[i\chi({\bf b},z_1,z_2)\right]$,  changes by 
\begin{equation}
\Delta \chi ({\bf b},z_1,z_2) = i{\sigma_{pN}\over 2}\int_{z_1}^{z_2} \rho({\bf r}) dz  - {1\over \hbar v}  \int_{z_1}^{z_2} U_R ({\bf r}) dz \label{Dchi}
\end{equation}
between two binary collisions happening between positions ${\bf r}_1 = ({\bf b},z_1)$ and   ${\bf r}_2 = ({\bf b},z_2)$. Between the two positions, the leading particle is assumed to move along a straight-line path with transverse coordinate ${\bf b}$. After one collision, the protons are allowed to move in a different direction. As long as their energies are large enough, they can be appropriately described by eikonal wavefunctions.  Their wavefunctions may be further influenced by a real mean-field potential $U_R$.
However, instead of amplitudes, as done in Ref. \cite{aumann:2013:PRC}, we will work with the probabilistic reaction propagator 
\begin{equation}
{\cal R}({\bf b},z_1,z_2)=1- |S({\bf b},z_1,z_2)|^2, \label{rb}
\end{equation} 
which measures the nucleon survival probability for an impact parameter ${\bf b}$. Therefore, within the present approximation, the model does not retain explicit sensitivity to the real part of the optical potential $U_R$.

In (p,3p) reactions, the reaction propagators is given by three factors: (a) The incoming proton propagates to position ${\bf r}_1 = ({\bf b},z_1)$, collides with another proton and, in their c.m., they scatter to angle $\Omega_1$. (b) One proton emerges out of the nucleus with impact parameter ${\bf b}_1'$. (c) The leading proton collides against second time at position   ${\bf r}_2 = ({\bf b}_1,z_2)$ and scatter in their c.m. to angle $\Omega_2$. (d) The two protons emerge out of the nucleus with impact parameters ${\bf b}_2$ and ${\bf b}_2'$. Thus, in this formalism,
  \begin{eqnarray} 
 K({\bf b}, {\bf r}_1,{\bf r}_2) &=& {\cal R} ({\bf b}, -\infty, z_1) |\psi_\alpha({\bf r}_1)|^2W \left( \Omega_{1} \right)  \ . \ {\cal R} ({\bf b}'_1,z_1,\infty) \nonumber \\
 &\times& {\cal R} ({\bf b}_1,z_1,z_2) |\psi_\beta({\bf r}_2)|^2 W\left( \Omega_{2} \right) \ . \ {\cal R} ({\bf b}'_2, z_2,\infty) \ . \  {\cal R} ({\bf b}_2, z_2,\infty) . \label{propag}
 \end{eqnarray}
 The wavefunctions squared, $|\psi_\alpha({\bf r})|^2$, account for the knockout of a proton from an orbital with quantum numbers defined by $\alpha$. For quasi-free elastic scattering of static protons in the nonrelativistic limit, the sum of the laboratory angles $\theta_{1L} + \theta_{2L}$ equals $90^\circ$, with relativistic and Fermi-motion effects leading to deviations from this value. Here we introduce the functions $W(\Omega)$ representing the likelihood for scattering at angles $\Omega$ in the c.m. of the colliding pair. In the case of elastic collisions, 
  \begin{equation} 
 W \left( \Omega \right)=  {1\over \sigma_{el}} {d\sigma_{el}\left(\Omega\right)\over d \Omega}, \label{Welast}
 \end{equation}
 with $d\sigma_{el}/d\Omega$  and $\sigma_{el}$ being the differential and total elastic scattering cross sections, without Coulomb interaction.  
 In summary (see Fig. \ref{fig1}), the impact parameters ${\bf b}_1$, and ${\bf b}_1'$, correspond to vector perpendicular distances between the nuclear center  and the velocity vector for the leading proton, and of the proton knocked out from the nucleus, after the first collision. A similar notation, with ${\bf b}_2$  and ${\bf b}_2'$, is used for the second collision.  The spectroscopic information about the removed protons are encoded in the wavefunctions $\psi_i({\bf r}_j)$.  
 
 This formalism could be easily extended to inelastic collisions such as pion production, by replacing in  $W(\Omega)$  the cross sections for pion production, or any other particle produced during the collisions.   For (p,2p) collisions, we will use a similar method, with 
   \begin{equation} 
 K({\bf b}, {\bf r}_1) = {\cal R} ({\bf b}, -\infty, z_1) |\psi_\alpha({\bf r}_1)|^2W \left( \Omega_{1} \right)  \ . \ {\cal R} ({\bf b}'_1,z_1,\infty) {\cal R} ({\bf b}_1,z_1,\infty)  . \label{propag2}
 \end{equation}
 One can also extend it to more  than two binary collisions. In contrast to the method employed in Ref. \cite{aumann:2013:PRC} for (p,2p) reactions which includes quantum interference,  Eqs. \eqref{propag} and \eqref{propag2} are purely based on probabilistic concepts and quantum mechanical interference is lost. 
 
Multiple collisions formalisms using eikonal waves have been developed  by many authors (see, e.g.,  \cite{PhysRev.103.443,PhysRev.184.1745,Faeldt1970_DeuteronNucleus,WALLACE1973190,PhysRevLett.32.911,RevModPhys.46.279,HufnerPRC12.1888}), although parallel straight-line impact parameters are invoked in the derivations. The model presented here is pure probabilistic, as are other successful models used, e.g., in nucleon-knockout reactions with heavy ion collisions \cite{Bertulani:92,Hencken:1996,Han03}. But it goes beyond by allowing different particle propagation directions after each collision. This idea has been previously applied for (p,2p) reactions in Ref. \cite{aumann:2013:PRC} although in a different way than it is carried out here because it was intended to describe (p,2p) and use it as a spectroscopic tool in the analysis of reactions involving specific nuclear states. In some sense, when extended to sequential binary collisions the model developed here is similar to intranuclear cascade models. To describe (p,3p) reactions we will  use Eq. \eqref{propag}. This requires that, knowing the scattering angle $\Omega$ in the c.m. of the colliding pp pair,  a kinematic transformation  to the nucleus frame of reference is needed to determine  the impact parameter of the leading proton and of the emerging one.  
To compute the angular functions $W(\Omega)$ we adopt the SAID model for nucleon-nucleon (NN) collisions.  It is a partial-wave analysis framework providing NN scattering amplitudes, phase shifts, and observables over a wide range of energies \cite{Arndt1987_NN_PWA,Arndt1994_SAID,Arndt2007_SAID,Workman2012_GWU}.  It is worthwhile mentioning that we did no include a medium modified pp  elastic scattering angular distributions. Angular distributions are less critical for inclusive cross sections, and medium effects mainly renormalize attenuation, not angular sampling.

The inclusive sequential (p,3p) cross section is obtained by summing over all possible single-particle states, and performing an integration over the collision coordinates and scattering angles. This is followed by an integration over the incoming proton position ${\bf b}$, the whole expression being
    \begin{equation} 
\sigma_{(p,3p)} = \sum_{\alpha,\beta}\int d^2b \, d^3r_1 \,  d^3r_{2}\, d\Omega_1 d\Omega_2 K_{\alpha,\beta}({\bf b},{\bf r}_1,{\bf r}_2). \label{sigma}
 \end{equation}
The cross section for (p,2p) collisions is a subset of this formalism,  meaning that 
     \begin{equation} 
\sigma_{(p,2p)} = \sum_{\alpha}\int d^2b \, d^3r_1 \,   d\Omega_1  K_{\alpha}({\bf b},{\bf r}_1), \label{sigma2}
 \end{equation}
 which involves only one single-particle state, $\alpha$.  
 
 We emphasize that the present calculation yields inclusive theoretical cross sections obtained by summing over all occupied proton single-particle orbitals. Removal of deeply bound protons may populate highly excited residual nuclei that subsequently undergo particle evaporation or fragmentation and therefore may not contribute to experimentally identified $(A-1)$ or $(A-2)$ residues. Such secondary decay effects are not included in the present framework. While this may affect the absolute normalization of the calculated cross sections, the relative systematics and their dependence on neutron-skin thickness are governed primarily by attenuation and surface geometry and are therefore expected to be robust.

To make calculations tractable, we assume  that the reaction is coplanar  which simplifies the numerics considerably. Coplanar geometry has been shown to be dominant in (p,2p) reactions \cite{Cha77}.  This implies a cylindrical symmetry, reducing the number of integrations because the aforementioned impact parameters $b_1$ and $b_2$ are determined once the scattering angles $\theta_1$ and $\theta_2$ are known.  The remaining integrations are done using a Monte-Carlo sampling over the impact parameter, and positions ${\bf r}_1$ and ${\bf r}_2$. It is worthwhile mentioning that this procedure is much simpler than that adopted in Ref.  \cite{aumann:2013:PRC}, which involves a more elaborated calculation with less constraining assumptions. A schematic view of a (p,3p) reaction is shown in Figure \ref{fig1}, where the angles are shown in the nucleus system. For quasi-free elastic scattering and static protons within the nuclei, the angles $\theta_{1L}$ and $\theta_{2L}$ are exactly equal to $90^\circ$ because of energy-momentum conservation laws. But they will deviate from this value if one accounts for nuclear Fermi motion or inelastic scattering, not considered in this work. Although the proton--proton elastic cross section is symmetric in the center-of-mass frame (after removal of Coulomb effects), transformation to the laboratory (nucleus) frame leads to a forward focusing of the outgoing protons at high energies.

The calculation of the single-particle states $\psi_\alpha$ is based on a one-body mean-field approximation, in which protons and neutrons move independently in a static spherical potential. The proton separation energy is approximated by the negative of the highest occupied proton single-particle energy. We use a usual WoodsSaxon form factor dependent on the distance $r$ from the nuclear center as  
\[f(r) = \frac{1}{\left[{1+\exp\!\left({r-R}/{a}\right)}\right]},\] 
with $R = r_0 A^{1/3}$ fm and $a=0.65$ fm.  Its derivative, ${df}/{dr}$
enters the spin-orbit interaction. For each orbital $\alpha = (n,\ell,j)$, the wavefunction is obtained by solving the  Schr\"odinger equation with the isovector + spin-orbit potential 
\[U_{\mathrm{cent}}^{(\tau)}(r)
+ \delta_{\tau,p} V_C(r)+ V_{so}^{(\tau)}(r)\langle \bm{\ell}\cdot\bm{s} \rangle\] 
with $\langle \bm{\ell}\cdot\bm{s} \rangle = \frac{1}{2}\big[j(j+1)-\ell(\ell+1)-\tfrac{3}{4}\big]$. 
The central WS potential is then 
\[U^{(p)}_{\mathrm{cent}}(r) = \big(V_0 - V_1\big)\,f(r)\ \ \ {\rm for \  protons, \ and} \ \ \ U^{(n)}_{\mathrm{cent}}(r) = \big(V_0 + V_1\big)\,f(r) \ \ \ {\rm for \ neutrons.}\]
The isoscalar depth \(V_0\) and isovector coefficient \(V_{1,\mathrm{coeff}}\) define $V_1 = V_{1,\mathrm{coeff}}\,\delta$, where $\delta = (N-Z)/A$. For \(N=Z\), one has \(\delta=0\) and the isovector contribution vanishes identically. Protons feel the Coulomb field of a uniformly charged sphere of radius \(R\), and the radial spin-orbit potential is $V_{so}^{(\tau)}(r) = -({r_0^2}/{2r})V_{so}^{(\tau)}{df}/{dr}$,
with isospin dependence 
\[V_{so}^{(p)} = V_{so0} - V_{so1}\,\delta, \ \ \  {\rm and} \ \ \  V_{so}^{(n)} = V_{so0} + V_{so1}\,\delta.\] 
We keep the values $V_0=-57.8$ MeV, $V_{so0}=-22.0$ MeV and $V_{so1}= 14.0$ MeV fixed. For $N\ne Z$ we vary the isovector coefficient $V_{1,\mathrm{coeff}}$ so that the magnitude of the  binding energy of the last filled orbital coincides with the proton separation energy of the nucleus. For $N=Z$ the scalar potential  $V_0$ is varied to reproduce the separation energy of the last orbital. The same procedure is applied to the potential for the neutrons.  The proton and neutron separation energies are obtained from the ENDF database \cite{ENDF}.

We use HFB densities for attenuation, and Woods-Saxon single-particle wave functions for the knockout calculations. This procedure renders calculations less complicated and may be justified since  the radial tails (important for skins) are dominated by density, while single-particle wave-function details are secondary for inclusive observables.

Although (p,2p) and (p,3p) reactions remove protons, their cross section depend on the total nuclear density profile  and therefore exhibits sensitivity to the neutron distribution and, in particular, to the neutron skin thickness. The sensitivity arises primarily from absorption in initial- and final-state interactions (ISI/FSI). Because $\sigma_{pn} \approx 2$--$3 \times \sigma_{pp}$ for energies in the range 200 MeV/nucleon -- 1 GeV/nucleon, attenuation of the incoming and outgoing fast protons depends more strongly on $\rho_n(r)$ than on $\rho_p(r)$. Consequently, the  cross sections and the transparencies may vary by 5--20\% or more between density models with different neutron skins. Ignoring the real potential $U_R$ in Eq. \eqref{Dchi}, the attenuation factors are determined by the phase-shifts between collisions through the relation
\begin{equation}
\Delta \chi ({\bf b},z_1,z_2) = i{\sigma_{pp}\over 2}\int_{z_1}^{z_2} \rho_p({\bf r}) dz  + i{\sigma_{pn}\over 2}\int_{z_1}^{z_2} \rho_n({\bf r}) dz,  \label{Dchi2}
\end{equation}
which explicitly shows the separate dependence on the proton $\rho_p$ and neutron density $\rho_n$. 
The incoming and outgoing protons have analogous factors  with one of the limits set to $z_i\to \pm\infty$, as explained previously. A thicker neutron skin decreases the transparency for peripheral incident proton, increasing $S({\bf b},-\infty, z_1)$ in Eq. \eqref{rb} and reducing the cross section. This argument, based on parallel straight-line trajectories for incoming and outgoing protons is often utilized to justify the high energy collisions as a probe of neutron skin. In our formalism, we allow the protons to scatter laterally and the effects of neutron skin could be more subtle, as we show below.

To calculate the cross sections we need  (a) the elastic proton-proton cross sections, (b) the nuclear ground state densities and (b) the (medium modified) nucleon-nucleon cross sections. As we explained earlier, the elastic pp cross sections in Eq. \eqref{Welast} are computed with the SAID model for nucleon-nucleon (NN) collisions \cite{Arndt1987_NN_PWA,Arndt1994_SAID,Arndt2007_SAID,Workman2012_GWU}.  The medium modified nucleon-nucleon cross sections at the different bombarding energies are calculated according to the model developed in Ref. \cite{BertulaniConti10}. The  nuclear densities are calculated using  the Hartree-Fock-Bogoliubov method with  22 Skyrme interactions denoted as SIII \cite{BEINER197529}, SKA and SKB \cite{KOHLER1976301},  SKM* \cite{BARTEL198279}, SKP \cite{DOBACZEWSKI1984103}, UNE0 and UNE1 \cite{STOITSOV20131592}, SKMP \cite{Bennour1989}, SKI2, SKI3, SKI4 and SKI5 \cite{REINHARD1995467}, SLY230A \cite{Chabanat:97}, SLY4, SLY5, SLY6, and SLY7 \cite{CHABANAT1998231},  SKX \cite{Brown:1998}, SKO \cite{Reinhard:PRC.60.014316},  SK255 and SK272 \cite{Agrawal2003}, HFB9 \cite{GORIELY2005425} and SKXS20 \cite{Dutra:PRC.85.035201}. We restrict our study to inclusive cross sections, where no definite knowledge of specific single-particle states are experimentally measured.

A mixed volume-surface pairing interaction was added with the form
\begin{equation}
v({\bf r},{\bf r}')= v_0 \left( 1- {1\over 2}  {\rho \over \rho_0}\right) \delta({\bf r}-{\bf r}'), \label{vpair}
\end{equation}
where $\rho(r) = \rho_n (r) + \rho_p(r)$ is the isoscalar local density, with the pairing strength being the same for neutrons and protons, and  a standard paring strength $v_0 = -258.2$ MeV \cite{Bertsch:PRC79.034306}. The saturation density is taken at $\rho_0 = 0.16$ fm$^{-3}$. The delta-function pairing force requires the use of a cutoff energy for the quasiparticles, chosen to be $E_{max} = 60$ MeV.   The HFB calculations were done with a modified version of the code HFBTO code \cite{STOITSOV20131592}.  

 \begin{figure}[t]
\begin{center}
\includegraphics[scale=0.37]{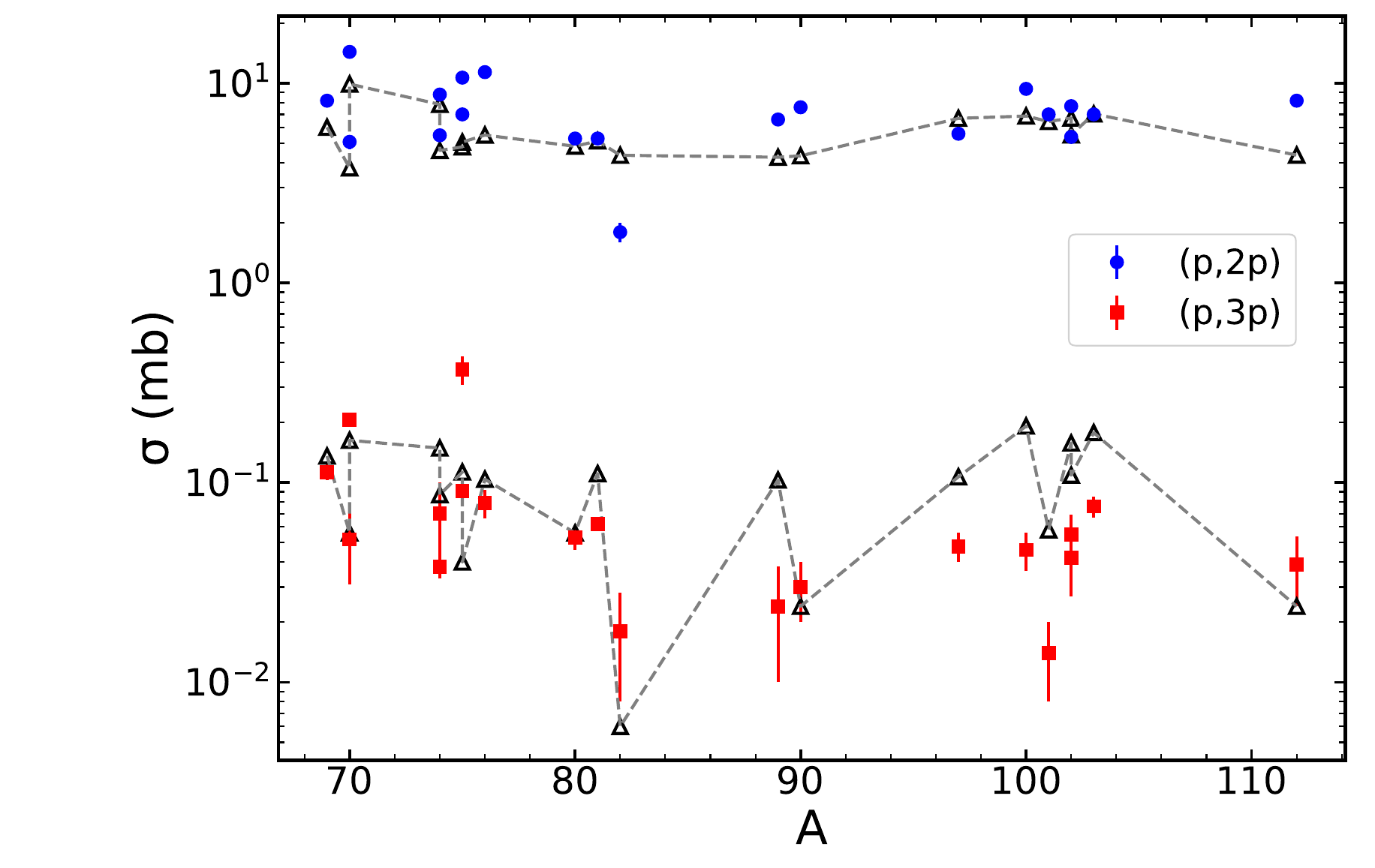}
\caption{ Experimental cross sections (filled symbols) \cite{PhysRevLett.125.012501} compared to our theoretical calculations (open diamonds) for (p,2p) and (p,3p) reactions as a function of the nuclear mass. The dashed lines are guides to the eye. \label{sigfig}}
\end{center}
\end{figure} 

\subsection{Numerical Results}

 Our results for the cross sections are shown in Fig. \ref{sigfig} (open diamonds) compared to the experimental data collected from Ref. \cite{PhysRevLett.125.012501} (filled symbols)  for (p,2p) and (p,3p) reactions as a function of the nuclear mass. The dashed lines are guides to the eye. Despite the logarithmic scale and far from perfect reproduction of the measured values, the calculations  follow the trend of the data with the nuclear mass dependence. In a few cases the calculations differ from the data by an order of magnitude. But in most cases the differences are about a factor 2-5.   The pronounced reductions of the $(p,3p)$ cross sections observed around $A=82$ and $A=101$ reflect changes in shell structure and surface density gradients along the isotopic chain. These structural effects modify the effective attenuation and spatial localization of the knockout probability, leading to enhanced suppression of the two-proton removal channel at specific neutron numbers.

 The level of agreement of our calculations with experimental data is encouraging, given the simplicity of the eikonal formalism and the assumption of sequential collisions and coplanar scattering. The coplanarity assumption is known to be well justified for (p,2p) collisions and has been shown to compare well with data \cite{Cowley:1995,PhysRevC.57.3185,Neveling:2002,PhysRevC.71.064606} and the interpretation for the good reasonably good agreement is because of a similar amount of absorption along the path of the protons in coplanar and non-coplanar trajectories. In other words, only for off-plane measurements of differential cross sections the non-coplanar feature is of relevance, while for total cross sections, the azimuthal directions of the protons average out. A similar argument may plausibly apply to (p,3p) reactions, although this assumption is less well tested.

Based on the relative success in describing the magnitude of the inclusive (p,2p) and (p,3p) cross sections within the sequential collision assumption, we now make an assessment of the dependence of  (p,2p) and (p,3p) reactions on the neutron skin  thickness $\Delta R_{np} $.  In order to test the dependence of (p,2p) and (p,3p) reactions on $\Delta R_{np}$, we consider a long Sn isotopic chain with neutron number varying within the range $N=50-88$.  

 \begin{figure}[t]
\begin{center}
\includegraphics[scale=0.42]{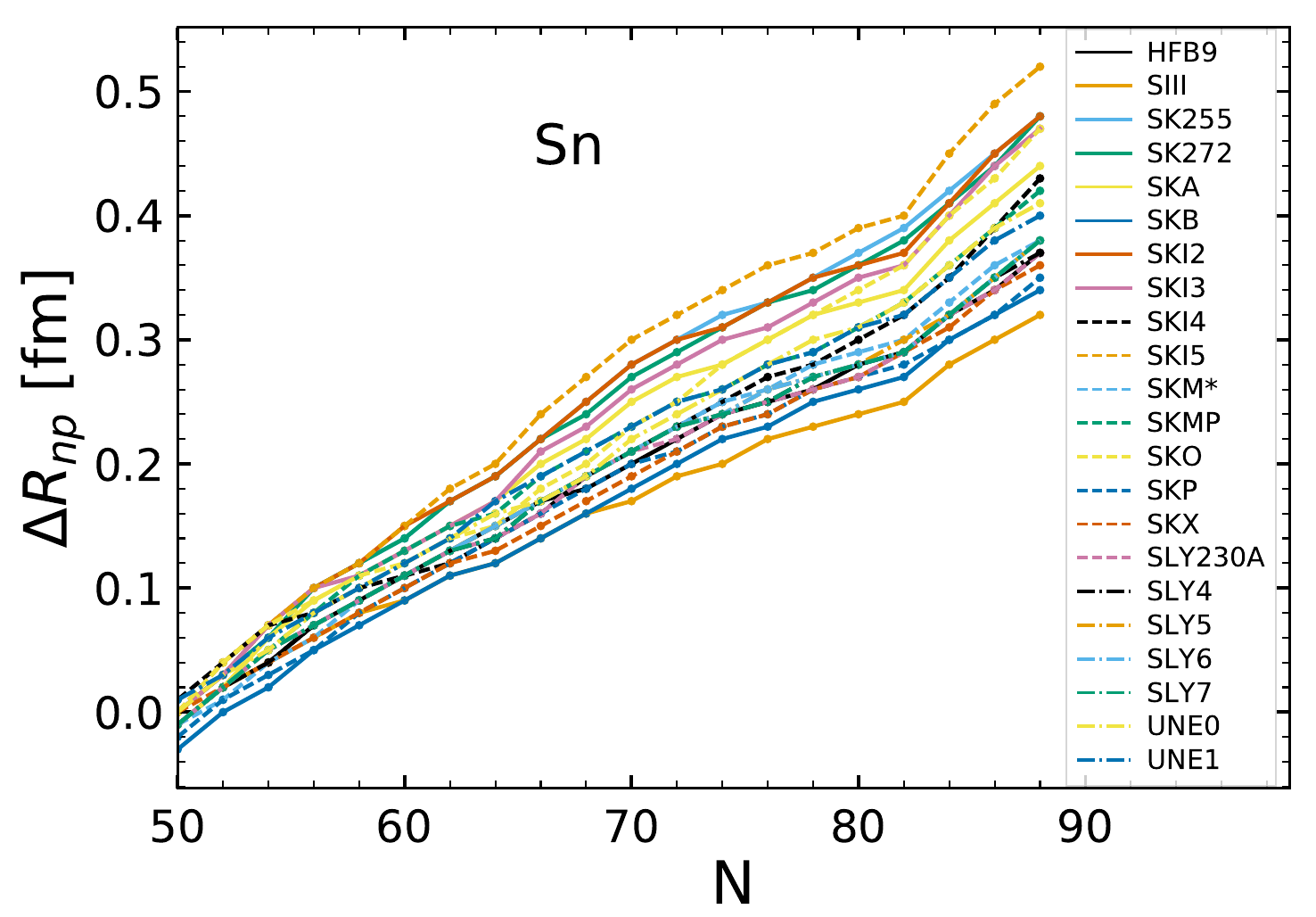}
\caption{ Neutron skin $\Delta R_{np} $ for Sn isotopes calculated with HFB using 22 Skyrme interactions listed in the text. Each one of the lines corresponds to one of the interactions and they are  guides
to the eye. \label{Sn-skin}}
\end{center}
\end{figure} 

In Fig. \ref{Sn-skin}, we plot the neutron skin calculated with the 22 Skyrme interactions listed previously.  The figure shows the evolution of the neutron skin thickness along the tin (Sn) isotopic chain as a function of neutron number  $N$. Each curve corresponds to a distinct Skyrme parametrization and illustrates the systematic theoretical spread associated with the isovector sector of the nuclear interaction.
A clear and robust trend emerges: the neutron skin thickness increases monotonically with neutron number for all interactions. This reflects the progressive excess of neutrons and the resulting outward extension of the neutron density relative to the proton density as neutrons are added to the system. For neutron-deficient isotopes near 
$N\sim Z=50$ the neutron  skin  is small and close to zero, while for neutron-rich isotopes approaching  $N\sim 90$, the neutron skin reaches values of about  $0.3-0.5$ fm, depending on the interaction.
Despite the common qualitative behavior, there is a sizable quantitative spread among the models, which increases with neutron number. This growing dispersion is a manifestation of the different density dependences of the symmetry energy encoded in the various Skyrme forces. Interactions with a stiffer symmetry energy generally predict larger neutron skins, especially for the more neutron-rich isotopes, while softer symmetry energies lead to systematically smaller values of  $\Delta R_{np}$.	
Overall, the figure highlights the sensitivity of neutron skin thickness in Sn isotopes to the isovector properties of the nuclear equation of state. The pronounced model dependence at large 
$N$ underscores the importance of experimental constraints on neutron skins in medium-to-heavy nuclei, as such data can significantly reduce uncertainties in the symmetry energy and improve the predictive power of nuclear energy-density functionals.

\begin{figure}[t]
\begin{center}
\includegraphics[scale=0.6]{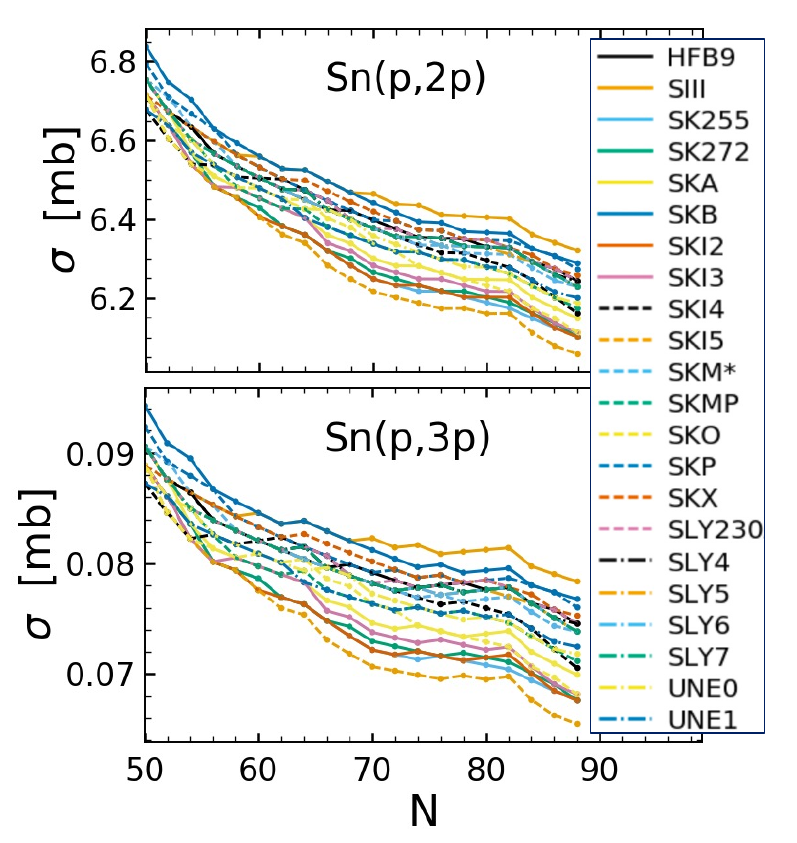}
\caption{ Calculated total cross sections for proton-induced one- and two-proton knockout reactions along the tin isotopic chain, shown as functions of neutron number 
$N$. The tin projectiles are assumed to have incident energy of 250 MeV/nucleon. The upper panel corresponds to the Sn(p,2p) reaction, while the lower panel shows results for the  Sn(p,3p) reaction. In both cases, the calculations are performed using HFB ground-state densities obtained with 22 different Skyrme energy-density functionals. \label{Sn-skin2}}
\end{center}
\end{figure} 

The two panels  in figure \ref{Sn-skin2} display calculated total cross sections for proton-induced one- and two-proton knockout reactions along the tin isotopic chain, shown as functions of neutron number 
$N$. The tin projectiles are assumed to have incident energy of 250 MeV/nucleon. The upper panel corresponds to the Sn(p,2p) reaction, while the lower panel shows results for the  Sn(p,3p) reaction. In both cases, the calculations are performed using HFB ground-state densities obtained with 22 different Skyrme energy-density functionals, allowing an assessment of model dependence associated with nuclear structure inputs.
A clear systematic trend is observed in both reactions: the cross sections decrease monotonically with increasing neutron number. As neutrons are added, the growing neutron excess leads to the development of a neutron skin, which effectively shields the protons in the nuclear interior. This reduces the probability for the incoming proton to interact with and remove one or two protons, resulting in smaller knockout cross sections for more neutron-rich isotopes. The overall reduction is moderate but robust across all interactions, indicating that this trend appears to be driven primarily by general geometric and isovector-density effects rather than by details of a specific parametrization..
The spread among the curves reflects the sensitivity of the calculated cross sections to the underlying Skyrme interactions. This model dependence increases slightly toward larger $N$, mirroring the growing dispersion seen in the predicted neutron skin thickness for neutron-rich Sn isotopes. Interactions that generate larger neutron skins tend, on average, to predict smaller (p,2p) and (p,3p) cross sections, consistent with increased attenuation of the incident proton flux by the neutron-rich surface region. The dependence on neutron skins is easily understandable if we use the simple argument arising from the Glauber method, as the variation $\delta \sigma \sim \delta \rho_n \sigma_{pn} dz$ which peaks at the surface and strongly depends on the $ \sigma_{pn}$ cross section.

Comparing the two panels, the relative sensitivity to neutron number is more pronounced for the 
(p,3p) reaction than for (p,2p). A quantitative estimate of the effect yields
\begin{equation}
\frac{d \ln \sigma_{p,3p}}{d \Delta R_{np}} \sim -(0.2-0.4) \ {\rm fm}^{-1}.
\end{equation}
This reflects the more exclusive nature of double-proton removal, which probes deeper into the nuclear interior and is therefore more strongly affected by changes in proton density distributions and by surface-to-volume competition induced by the neutron skin. Consequently, (p,3p) reactions may offer enhanced sensitivity to isovector nuclear structure effects and could serve as a complementary indicator of neutron-skin systematics and the density dependence of the symmetry energy. 

Semiclassical and eikonal descriptions of multistep nucleon-induced reactions have a long history. Early work by Kawai and Weidenm\"uller \cite{KawaiWeidem} developed a semiclassical treatment of two-step nucleon-nucleus processes that is closely related in spirit to the probabilistic multistep approach adopted here. More recently, an eikonal description of (p,3p) reactions was presented by G\'omez-Ramos  in Ref. \cite{Ramos2024}. While both approaches employ eikonal dynamics, the present framework emphasizes inclusive observables and density-driven attenuation effects rather than exclusive final-state amplitudes. Benchmarks for (p,pN) reactions have also been studied in Refs. \cite{RDM2020,YoshRam2018}.

\section{Momentum distribution formalism}
\label{sec:Mformalism}

A description of the angular and momentum distribution of the recoiled nucleus, such as that published in Ref. \cite{aumann:2013:PRC} for (p,2p) reactions, is not appropriate in the case of (p,3p) reactions. We develop here a different approach.  The purpose of the calculation is to determine the longitudinal momentum
dispersion of  projectile fragments produced in sequential proton-proton collisions, for (a) one-proton removal, $(p,2p)$, leaving a residue $(Z-1,N)$, and (b) two-proton removal, $(p,3p)$, leaving a residue $ (Z-2,N)$ from a tin projectile of mass number $A$ incident on a proton target at 250~MeV, same case as in Ref. \cite{PhysRevLett.125.012501}. The observable of interest is the longitudinal (beam-axis) momentum width
$\Delta_\parallel(A) = \sqrt{ \langle P_z^2 \rangle - \langle P_z \rangle^2 },$ where $P_z$ is the recoil momentum of the primary fragment. The calculation is strictly limited to sequential collisions with evaporation and ablation explicitly excluded. We only assess the effects imprinted on longitudinal momentum distributions because transverse momentum distributions are known to be more sensitive to additional details of the nucleon-nucleon interactions \cite{AumBar21}.

In the present implementation, the proton density is modeled by HFB + Skyrme interactions. For simplicity, we only show the calculations for the SLy4 force, as the spread of the results over the different forces is similar to what is shown in Fig. \ref{Sn-skin2},  increasing   with neutron number along the tin isotopic chain. The intrinsic momentum of each removed nucleon is sampled explicitly from a local Fermi sea determined by the local density. In the local density approximation, the Fermi momentum (in units of MeV/$c$) of protons is
\begin{equation}
p_{F}(r)=\hbar c\,
\bigl[3\pi^2 \rho_p(r)\bigr]^{1/3},
\label{eq:pF}
\end{equation}
with $\rho$ given in unites of fm$^{-3}$.
The local momentum variance is $\left<p^2\right>(r)= 3p_{F}^2(r)/5$.
The corresponding momentum distribution is assumed to be isotropic and uniform within the Fermi sphere,
\begin{equation}
f_p(\mathbf{p};r)=\frac{3}{4\pi p_{F}^3(r)}\,
\Theta\!\bigl(p_{F}(r)-|\mathbf{p}|\bigr).
\end{equation}
For each Monte Carlo event, the spatial position of a removed nucleon is sampled with weight
\begin{eqnarray}
w(\mathbf{r}_1,b)
&\propto&
\rho_p(\mathbf{r})\,
 \sum_{\alpha,\beta} K_{\alpha}({\bf b},{\bf r}_1), \nonumber \\
w(\mathbf{r}_1;\mathbf{r}_2,b)
&\propto&
\rho_p(\mathbf{r})\,
 \sum_{\alpha,\beta} K_{\alpha,\beta}({\bf b},{\bf r}_1,{\bf r}_2),
\end{eqnarray}
using the functions  $ K_{\alpha}({\bf b},{\bf r}_1)$ and $K_{\alpha,\beta}({\bf b},{\bf r}_1,{\bf r}_2)$ described in Eqs. \eqref{propag} and \eqref{propag2}.
After that a momentum vector ${\bf p}_i$ is drawn from the corresponding local Fermi sphere at each point ${\bf r}_j$ ($j=1$ or $j=1,2$). Here $\mathbf{r}_1$ and $\mathbf{r}_2$ denote the spatial coordinates of the first and second knocked-out protons, respectively, measured from the center of the projectile nucleus, and $\mathbf{r}\equiv(\mathbf{r}_1,\mathbf{r}_2)$ is used as a shorthand notation in the (p,3p) case. The recoil momentum of the primary fragment in (p,2p) reactions is given by 
\begin{equation}
\mathbf{P}_{\rm frag} ({\bf r}_1) =-\sum_{i=1}^{Z-1} \mathbf{p}_i. \label{pfrag}
\end{equation} 
For (p,3p) reactions another sum is added to $\mathbf{P}_{\rm frag}$ at ${\bf r}_2$  with the upper limit of the sum over ${\bf p}_i$ set to $Z-2$.  Then the fragment momentum $\mathbf{P}_{\rm frag}$ is projected along the longitudinal (beam-axis) direction, to obtain $P_Z= {P}_{\rm frag} \cos (\theta)$. 

A mean value of the fragment momentum squared is calculated from 
\begin{eqnarray}
\left<P_{Z}^2\right>_{p,2p}(b) &=& \frac{\int d^3r_1 w(\mathbf{r}_1,b) \left<P_Z^2\right>(\mathbf{r}_1)}{\int d^3r_1 w(\mathbf{r}_1,b) }, \nonumber \\
 \left<P_{Z}^2\right>_{p,3p}(b) &=& \frac{\int d^3r_1 d^3r_2 w(\mathbf{r}_1,\mathbf{r}_2,b) \left<P_Z^2\right>(\mathbf{r}_1,\mathbf{r}_2)}{\int d^3r_1 d^3r_2 w(\mathbf{r}_1,\mathbf{r}_2,b) }, \label{PZ}
\end{eqnarray}
and similar averages are used for  $\left<P_{Z}\right>$.
Integration over impact parameters with weight $2\pi b\,db\,P_{Z}^n(b)$, with $n=1,2$, yields the full momentum distribution and the mean value squared of the primary fragment. The longitudinal momentum dispersion is extracted as
\begin{equation}
\Delta_\parallel
=
\sqrt{\langle P_z^2\rangle - \langle P_z\rangle^2}.
\end{equation}

\subsection{Numerical Results}

\begin{figure}[t]
\begin{center}
\includegraphics[scale=0.5]{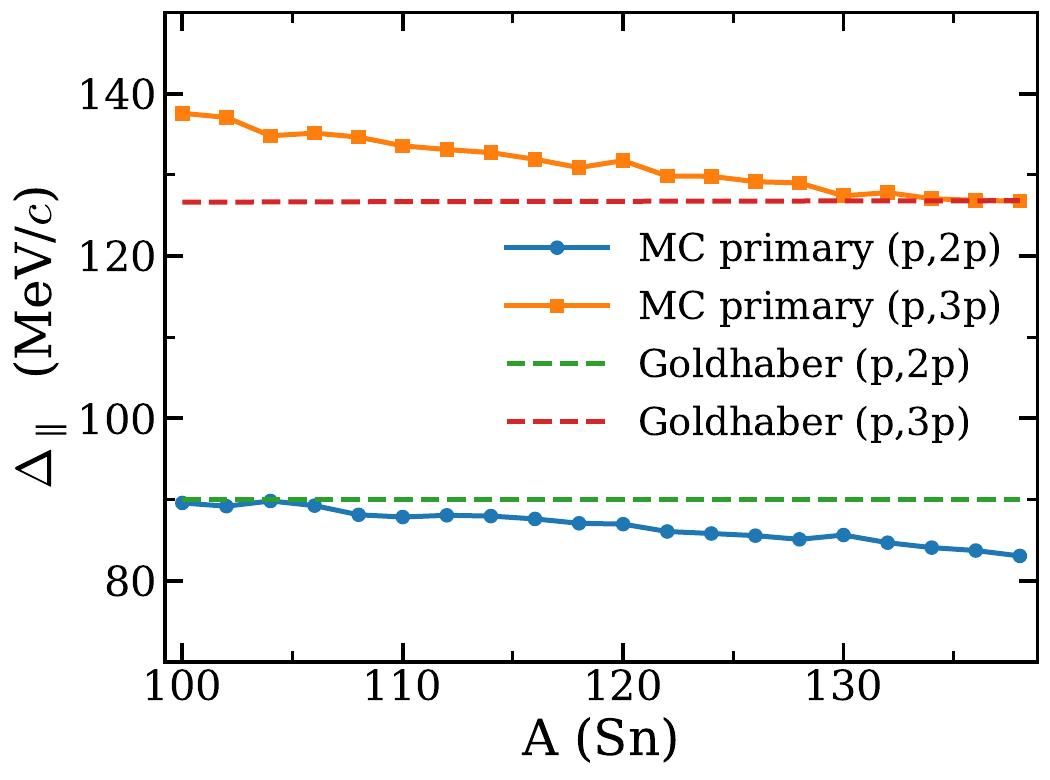}
\caption{ Longitudinal momentum dispersion $\Delta_{\|}$  of primary projectile-like fragments produced in (p,2p) (blue circles) and  (p,3p) (orange squares) reactions on Sn isotopes at 250 MeV, calculated with the vectorized Monte Carlo abrasion model. The dashed horizontal lines show the corresponding Goldhaber model predictions with a fixed width parameter $p_0 = 90$ MeV/$c$. The Monte Carlo results exhibit a clear mass dependence, reflecting surface-dependence of proton removal and the reduction of the local Fermi momentum in neutron-rich nuclei, an effect absent in the Goldhaber approach. \label{momdis}}
\end{center}
\end{figure} 

Figure~\ref{momdis} displays the longitudinal momentum dispersion
\(\Delta_\parallel\) of primary projectile-like fragments produced in
\((p,2p)\) and \((p,3p)\) reactions on Sn isotopes at an incident proton energy
of 250~MeV. The results obtained with the present vectorized Monte Carlo (MC)
model are compared to the corresponding Goldhaber model predictions,
shown as horizontal dashed lines. Several salient physical features emerge from
this comparison, which we discuss in detail below. In the Goldhaber model, the longitudinal momentum width is also computed using
\cite{Goldhaber1974},
\begin{equation}
\Delta_{\|} = p_0 \sqrt{ \frac{A_f(A_p-A_f)}{A_p-1}}, \label{Goldh}
\end{equation}
where $A_f=A-1$ for (p,2p) and $A_f = A-2$ for (p,3p) reactions. 
Our model provides a microscopically motivated extension of the physical ideas underlying the Goldhaber model \cite{Goldhaber1974,PhysRevC.39.460}. In Eq. \eqref{Goldh}, $p_0$ is the Goldhaber width parameter characterizing the effective Fermi momentum of nucleons in the statistical fragmentation model. Throughout this work we adopt the standard value $p_0=90$~MeV/$c$, commonly used in analyses of relativistic fragmentation data. In the latter, the momentum width arises from the vector sum of randomly oriented Fermi momenta, yielding $\Delta_\parallel^2 \propto \rho\, (3p_F^2/5)$, with a constant bulk Fermi momentum $p_F$. In contrast, the present calculation naturally incorporates the surface bias of proton removal reactions and the reduction of the local Fermi momentum in low-density regions.
Although only protons are removed, the neutron distribution modifies the geometry of the interaction region and shifts the sampling of proton momenta toward regions of lower density. This results in a systematic reduction of $\Delta_\parallel$ with increasing neutron skin thickness within the present reaction framework.  The effect is more pronounced for the $(p,3p)$ channel, which samples the surface more strongly and involves a larger recoil.

A clear hierarchy between the two reaction channels is observed,
\begin{equation}
\Delta_\parallel^{(p,3p)} > \Delta_\parallel^{(p,2p)},
\end{equation}
over the entire isotopic chain. This ordering is physically expected and follows directly from the recoil mechanism of the primary fragment \cite{Tostevin1999}. As shown in Eq.~\eqref{pfrag}, the fragment momentum is given by the negative vector sum of the intrinsic momenta of the removed nucleons. For single-proton removal, the recoil momentum reflects the sampling of one Fermi momentum, whereas for two-proton removal the variance is increased by the incoherent sum of two independent nucleon momenta. This geometric effect leads to a gradual suppression of \(\Delta_\parallel\) with increasing neutron excess, amounting to a reduction of approximately \(5\!-\!7\%\) across the isotopic chain shown in Fig.~\ref{momdis}. Such an effect may be within the reach of modern experimental momentum resolution and constitutes a genuine structural signal. The reduction of the cross sections reported here arises from geometric and density-dependent effects and does not require explicit short-range correlations. We expect that short-range correlations would mostly affect high-momentum tails, not total cross sections, or the low momentum part of the spectrum.

While this qualitative behavior is captured by both the MC calculation and the Goldhaber model, the absolute magnitude of the dispersions differs significantly between the two approaches. The MC results yield typical values of \(\Delta_\parallel \simeq 85\!-\!90\)~MeV/\(c\) for \((p,2p)\) and \(\Delta_\parallel \simeq 125\!-\!135\)~MeV/\(c\) for \((p,3p)\), whereas the Goldhaber model produces fixed, $A$-independent values determined solely by the parameter \(p_0\). This discrepancy reflects the fundamentally different physical assumptions underlying the two descriptions. Because the Goldhaber model lacks any spatial or density dependence, it cannot reproduce the observed mass dependence or the sensitivity to neutron skins. Notice that the fact that the  results of the Goldhaber model approximately agree with that of (p,2p) reactions at small $N$, and with that of (p,3p) reactions at large $N$, is purely accidental. Had we used a value different from $p_0=90$ MeV/$c$ for the parameter in Eq.  \eqref{Goldh}, we would obtain a different set of constants for  $\Delta_{\|}$ in the model.

The comparison between the vectorized Monte-Carlo knockout calculation and the Goldhaber model highlights the limitations of a purely statistical description of fragment recoil. The Goldhaber formula assumes removal from a uniform Fermi gas and predicts a simple scaling of the longitudinal width with the final fragment mass. By contrast, the Monte-Carlo calculation explicitly accounts for the spatial dependence of the removal probability and the local Fermi momentum.

Benchmark comparisons of calculated longitudinal momentum widths with experimental data for light nuclei such as $^{12}$C and $^{16}$O, where the number of active orbitals is small, would provide a valuable test of the present framework. Such an analysis is beyond the scope of the present work and will be addressed in future studies.

\section{Conclusions}
\label{sec:conclusions}

The results presented in this work indicate that inclusive (p,2p) and (p,3p) reactions exhibit a systematic sensitivity to neutron-skin thickness through both total cross sections and longitudinal momentum distributions. The decrease of cross sections with increasing neutron excess observed along the tin isotopic chain arises primarily from enhanced attenuation of the incoming and outgoing protons by the neutron-rich surface region. Because neutron-proton cross sections exceed proton-proton cross sections at intermediate and relativistic energies, the neutron density plays a dominant role in shaping the reaction transparency. As a result, nuclei with thicker neutron skins exhibit reduced effective access to interior protons, leading to smaller knockout cross sections. We have clarified that while (p,2p) reactions do probe the nuclear interior, strong attenuation leads to effective surface weighting, particularly in neutron-rich nuclei. This clarification reconciles the present discussion with Refs. \cite{aumann:2013:PRC,Bertulani.104.L061602}.

This effect is amplified in two-proton removal reactions. The $(p,3p)$ channel probes deeper into the nucleus and samples the surface more strongly than $(p,2p)$, making it particularly sensitive to changes in the surface-to-volume ratio induced by neutron skins. Consequently, $(p,3p)$ cross sections decrease more rapidly with neutron number, and their variation across different Skyrme energy-density functionals is more pronounced. This enhanced sensitivity suggests that systematic measurements of (p,3p) reactions along isotopic chains could provide valuable complementary constraints on neutron-skin thickness and, by extension, on the density dependence of the symmetry energy.

We have developed a microscopic Monte Carlo model for the longitudinal momentum dispersion of primary fragments in proton-induced knockout reactions. The model incorporates realistic geometry, channel selection, and local Fermi motion, and allows a direct investigation of neutron skin effects along isotopic chains. The results indicate that $(p,2p)$ and $(p,3p)$ momentum distributions are also sensitive to surface properties of neutron-rich nuclei, with enhanced sensitivity in the two-proton removal channel. But we emphasize that the present formalism neglects quantum interference effects between competing reaction paths; while this approximation is justified for inclusive observables, it may limit its applicability to exclusive or polarization-dependent measurements. Interference effects are known to affect asymmetries and widths in (p,2p) momentum distributions when specific single-particle states are probed \cite{AumBar21} .

The results shown in Fig.~\ref{momdis} indicate that longitudinal momentum dispersions of primary fragments encode nontrivial information on nuclear surface structure. In particular, they provide access to density-dependent Fermi motion and neutron skin effects that are completely washed out in simplified models. When combined with microscopic densities from Hartree-Fock-Bogoliubov calculations employing Skyrme functionals, systematic measurements of \(\Delta_\parallel\) in \((p,2p)\) and \((p,3p)\) reactions along isotopic chains offer a promising new avenue to constrain the isovector sector of the nuclear energy density functional. Two-proton removal involves multiple attenuation paths and deeper penetration into the nuclear surface region. As a result, neutron-rich surface layers suppress (p,3p) yields more strongly than (p,2p), amplifying sensitivity to $\Delta R_{np}$ and to the density dependence of the symmetry energy. Future extensions include the incorporation of quantum interference effects, secondary de-excitation, short-range correlations. On the experimental side, systematic measurements at facilities such as FRIB and FAIR will enable stringent tests of these predictions and offer new opportunities to constrain the isovector sector of the nuclear energy density functional.

The observed trends can be qualitatively and semi-quantitatively connected to the density dependence of the nuclear symmetry energy. In self-consistent energy density functional theories, the neutron skin thickness of heavy nuclei correlates strongly with the symmetry-energy slope parameter \(L\). Skyrme parameterizations with larger \(L\) values predict thicker neutron skins and more diffuse neutron surfaces. The  correlation of the cross sections with $\Delta R_{np}$ is also similar to that found between $\Delta R_{np}$ vs. $L$ (see, e.g.,  Ref.  \cite{Centelles2009,Bertulani.PRC.100.015802}. Within the present framework, an increase in \(L\) implies a larger \(\Delta R_{np}\), which shifts the proton knockout region to lower densities and reduces the effective Fermi momentum entering Eq.~(\ref{PZ}). As a result, one expects a monotonic decrease of \(\Delta_\parallel\) with increasing \(L\). For typical Skyrme functionals spanning \(L\simeq 30\!-\!100\)~MeV, the associated variation in neutron skin thickness for Sn isotopes is of order \(0.1\!-\!0.3\)~fm, which translates into a few MeV/\(c\) change in \(\Delta_\parallel\) in the present calculations. This sensitivity is comparable to that of interaction cross sections and complementary to electroweak probes such as parity-violating electron scattering \cite{Horowitz2014}. But we emphasize that the present results are obtained within a probabilistic, incoherent multistep framework and are intended to highlight robust qualitative trends rather than provide precision extractions of neutron skin thickness or symmetry-energy parameters.

\bigskip

{\bf Acknowledgments}

C.A.B. is supported by U.S. Department of Energy Office of Nuclear Physics under Contract No. DE-SC0026074 with East Texas A\&M University. R.V.L. is supported by INCT-FNA (Instituto Nacional de Ci\^encia e Tecnologia, F\'\i sica Nuclear e Aplica\c c\~oes), research Project No. 464898/2014-5, and also thanks to CAPES/CNPq.


\end{document}